\documentclass[doublecol]{epl2}

\title{Evidence of fractional matching states in nanoperforated Nb thin film grown on porous silicon}
\shorttitle{Evidence of fractional matching states in Nb thin film
grown on porous silicon}

\author{M. Trezza\inst{1} \and C. Cirillo\inst{1} \and S. L. Prischepa\inst{2} \and C.Attanasio\inst{1,3}}
\shortauthor{M. Trezza
 \etal}

\institute{
  \inst{1} Laboratorio Regionale SuperMat, CNR-INFM Salerno and
Dipartimento di Fisica \lq\lq E. R. Caianiello\rq\rq,
Universit\`{a} degli Studi di Salerno, Baronissi (Sa) I-84081,
Italy\\
  \inst{2} State University of Informatics and RadioElectronics, P. Brovka street 6, Minsk 220013, Belarus\\
  \inst{3} NANO\_ MATES,  Research Centre for NANOMAterials and
nanoTEchnology at Salerno University, Universit\`{a} degli Studi
di Salerno, Fisciano, (Sa) I-84084, Italy}

\pacs{74.25.Fy}{Transport properties of superconductors}
\pacs{74.25.Qt}{Vortex lattices, flux pinning, flux creep}
\pacs{81.05.Rm}{Porous materials}

\abstract{Resistive transitions have been measured on a perforated
Nb thin film with a lattice of holes with the period of the order
of ten nanometers. Bumps in the $dR$/$dH$ versus $H$ curves have
been observed at the first matching field and its fractional
values, 1/4, 1/9 and 1/16. This effect has been related to
different vortex lattice configurations made available by the
underlying lattice of holes.}

\begin{document}

\maketitle

\section{Introduction}

The nucleation of Abrikosov vortices \cite{Abrikosov57} in the
mixed state of type-II superconductors with periodic artificial
pinning centers attracted a great attention since 1970's. Recent
progress in the fabrication of nanostructures provides the
possibility to realize superconducting thin films containing
artificial defects as pinning sites with well-defined size,
geometry and spatial arrangement \cite{Mosh98,Schuller00}. Vortex
pinning was extensively explored by many groups to develop a
fundamental understanding of flux dynamics and for its relevance
in applications which require enhancements of the critical current
density. Thus, several types of artificial pinning centers, such
as square, rectangular or triangular arrays, have been introduced
in a controlled way in the superconducting films. In particular,
the use of regular array of pinning centers such as antidots
\cite{Mosh98,Fiory78,Lykov93,Castellanos97} or magnetic dots
\cite{Schuller00,Schuller97,Mosh99,Schuller08} brings to new
commensurability effects, which give additional insight into the
pinning properties of vortices. The most notable phenomenon for
these studies is the so-called matching effect which occurs when
the vortex lattice is commensurate with the periodic pinning
array. This situation occurs, in particular, at fractional or
integer values of the so-called first matching field
$H_1=\Phi_0/a_{0}^{2}$, i.e., when the applied field $H$
corresponds to one flux quantum, $\Phi_0=h/2e$, per unit cell
area, $a_{0}^{2}$, of the pinning array. Here $a_{0}$ is the
lattice constant of the pinning arrangement. As a result, at the
matching field, the critical current density, $J_c$, is
drastically enhanced \cite{Schuller00,Fiory78,Silhanek05} and
moreover, as a consequence of the Little-Parks effect
\cite{Little-Parks}, the upper critical magnetic field is
increased at the matching values. Recently antidot arrangements
with a big variety of symmetries have been investigated. Matching
effects have been reported in perforated Nb thin films for
antidots lattices with short range order \cite{Ziemann}, or
quasiperiodic fivefold Penrose structures \cite{Kemmler06}.
Moreover asymmetric pinning arrays have been suggested as
superconducting rectifiers \cite{Mosh05}.\\
If the artificial structure of defects is created by lithographic
technique, the matching fields are usually in the range of a few
oersteds. For this reason, matching effects are observed in a very
narrow temperature region, close to the critical temperature
$T_c$, for a reduced value $t=T/T_c\geq0.95$. In order to both
increase the matching field and decrease the temperature where the
effect is present, the period of the pinning structure should be
reduced to less than 100 nm. This gives, in fact, the possibility
to increase $H_1$ up to 1 tesla or even higher. A reasonable
method to achieve this goal is to use self-assembled substrates,
such as, for example, $Al_{2}O_{3}$ templates with characteristic
features in the nanometric scale \cite{Mosh07}. The pore diameter
in $Al_{2}O_{3}$ substrates could easily be varied in the range
25-200 nm with porosity (i.e. interpore spacing) around 50\%, and
this gives the possibility to achieve matching fields of thousands
of oersteds \cite{Mosh07}. To prepare $Al_{2}O_{3}$ substrates
bulk Al \cite{Prischepa-Cryogenics94}, Al foils \cite{Welp-PRB},
and deposited thick Al films were used \cite{Mosh07}.\\
Very recently, another very promising material for self-assembled
substrates and an optimum candidate for the Nb growth was
proposed, namely, porous silicon (PS) \cite{Trezza-JAP08}. PS is
constituted by a network of pores immersed in a nanocrystalline
matrix \cite{Pavesi} and it is a material which offers a
considerable technological interest in different fields, as for
instance micro and optoelectronics \cite{Collins} and gas sensing
\cite{Cheraga,Lysenko}. The diameter of pores, ${\O}$, in PS can
easily be varied from 200 nm down to 5 nm by using substrates with
appropriate doping (n or p) and different regimes of anodization.
The porosity, in fact, can be varied in the range 30-90\% by
adjusting parameters such as the acid solution, the anodizing
current density and the illumination of the substrate during the
anodization. The regularity of the pores arrangement, however, is
of the order of 10\% lower than the one observed in $Al_{2}O_{3}$
templates obtained by electrochemical oxidation \cite{Piraux}. It
has been demonstrated \cite{Trezza-JAP08} that thin Nb films
deposited on PS substrates can inherit their structure. The
resulting samples then consist of porous Nb thin films with in
plane geometrical dimensions, $a_{0}$ and ${\O}$, comparable with
the superconducting coherence length, $\xi(T)$. In these samples,
matching fields of the order of 1 Tesla
were experimentally observed \cite{Trezza-JAP08}.\\
Aim of this work is to deepen the study of the matching effect in
superconducting Nb thin films deposited on PS. Superconducting
properties  were investigated by transport measurements in the
presence of magnetic fields applied perpendicularly to the samples
surface, down to t = 0.52. As a consequence of the high density of
the pore network, the (H,T) phase diagram presents a deviation
from the classic linear dependence. This effect appears at the
matching field $H_{1}\approx$ 1 Tesla, a value larger than those
typical of periodic pinning arrays obtained both by lithographic
techniques and by using another kind of self-organized templates.
Moreover a new effect related to the commensurability between the
vortex lattice and the underlaying pinning structure was found. It
consists in the appearance of pronounced structures in the
derivative of the $R(H)$ curves, $dR$/$dH$, which can be observed
in correspondence of the first matching field and its fractional
values.

\section{Fabrication}

Porous layers were fabricated by electrochemical anodic etching of
n-type, 0.01 $\Omega$cm, monocrystalline silicon wafers. The
electrochemical dissolution was performed in 48\% water solution
of HF, applying a current density of 20 mA/cm$^{2}$. The
anodization time was chosen in the range of 0.5 - 4 min in order
to get porous layers with a thickness ranging from 0.5 to 4
$\mu$m. The pores extend on a surface of about 1 cm$^{2}$. The
integral porosity was estimated by gravimetry to be of about 50\%
\cite{Lazarouk}. The resulting porous substrates have ${\O}$=10 nm
and $a_{0}$ = 40 nm. For this lattice, if the formula
$H_1=\Phi_0/a_{0}^{2}$ for the square lattice is used, the
expected first matching field is $H_{1}$ = 1.3 Tesla.\\
Nb thin films were grown on top of the porous Si substrates in a
UHV dc diode magnetron sputtering system with a base pressure in
the low $10^{-8}$ mbar regime and sputtering Argon pressure of
$3.5\times10^{-3}$ mbar. In order to reduce the possible
contamination of the porous templates, the substrates were heated
at $120^{\circ}$C for one hour in the UHV chamber. The deposition
was then realized at room temperature after the cool off of the
substrates. Films were deposited at typical rates of 0.33 nm/s,
controlled by a quartz crystal monitor calibrated by low-angle
reflectivity measurements. Since the effect of the periodic
template would be reduced when the film thickness, $d_{Nb}$,
exceeds the pore diameter, ${\O}$, \cite{Trezza-JAP08} the Nb
thickness was chosen to be 8.5 nm for the sample analyzed in this
paper. A reference Nb thin film of the same thickness was grown on
a non-porous Si substrate in the same deposition run.

\section{Experimental results and discussion}

\begin{figure}
\includegraphics[width=8cm]{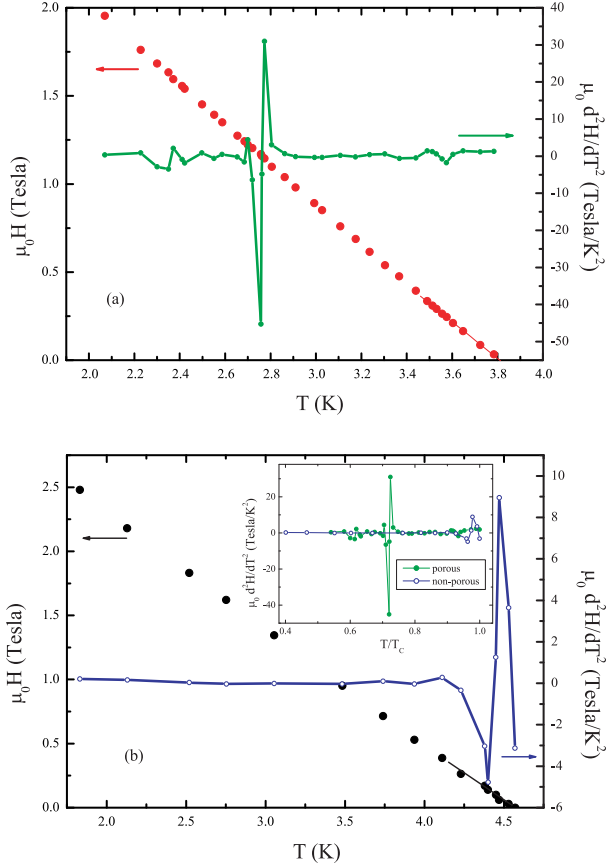} \caption{Left scale: Perpendicular upper
critical field $H_{c2\bot}$ vs. temperature of the Nb thin film
with $d_{Nb}$ = 8.5 nm grown on (a) porous template and (b)
non-porous reference substrate. The linear fits to the data close
to $T_{c}$ are also shown. Right scale: $dH_{c2\bot}^{2}$/$dT^{2}$
versus temperature. The inset shows the comparison between the
second derivatives as functions of the reduced temperature of two
samples, grown on the porous template (full circles) and on the
non-porous template (open circles). (Color online).} \label{Fig.1}
\end{figure}

The superconducting properties were resistively measured in a
$^{4}$He cryostat using a standard dc four-probe technique on
unstructured samples. The critical temperature was defined at the
midpoint of the $R(T)$ transition curves. The value of the
transition temperatures of the film grown on the porous substrate
and of the reference sample in the absence of the magnetic field
were $T_c$ = 3.83 K and $T_c$ = 4.53 K, respectively. The critical
temperature depression in the case of the porous sample is
consistent with what already reported in literature for  films
grown both on $Al_{2}O_{3}$ \cite{Mosh07} and on PS
\cite{Trezza-JAP08}. The first step for the characterization of
the behavior of the porous Nb sample in the presence of
perpendicular magnetic field is the determination of its ($H$,$T$)
phase diagram. The temperature dependence of the perpendicular
upper critical field, $H_{c2\bot}$, was obtained performing
resistance vs. field, $R(H)$, measurements at fixed values of the
temperature with a temperature stability of 1 mK.
$H_{c2\bot}$ was defined at the midpoint of each of the $R(H)$ curves.\\
In Fig. \ref{Fig.1} the ($H$,$T$) phase diagrams of the Nb thin
films are shown. In general, the perpendicular upper critical
field of superconducting films of thickness $\textit{d}$ obeys a
linear temperature dependence, $H_{c2\bot}(T)$ =
($\Phi_{0}$/2$\pi\xi_{0\parallel}^{2}$)(1-$T$/$T_{c}$)
\cite{Tinkham}. $\xi_{0\parallel}$ is the Ginzburg-Landau
coherence length parallel to the sample surface at $T$ = 0. The
temperature dependence of $\xi_{\parallel}$ is
$\xi_{\parallel}(T)$ = $\xi_{0\parallel}$/$\sqrt{1-T/T_{c}}$.
Another superconducting parameter to be taken into account is the
magnetic field penetration depth, $\lambda$, whose temperature
dependence is $\lambda(T)$ = $\lambda_{0}$/$\sqrt{1-T/T_{c}}$,
where $\lambda_{0}$ is the penetration depth at $T$ =0.

\begin{figure}
\includegraphics[width=8cm]{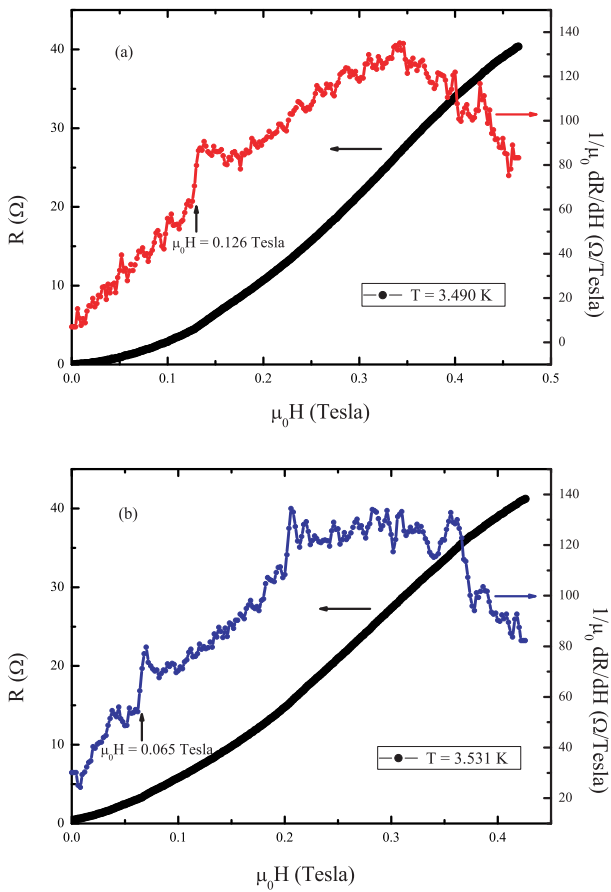} \caption{Left scale: $R(H)$ measurement
at (a) $T$ = 3.490 K and (b) $T$ = 3.531 K. Right scale: $dR$/$dH$
versus the applied magnetic field. In both panels the arrow
indicates the field where the bump is present. (Color online)}
\label{Fig.2}
\end{figure}

The $H_{c2\bot}(T)$ curve obtained for the Nb film deposited on
porous Si template, reported in Fig. \ref{Fig.1}(a), presents some
peculiarities, which indicate that the superconducting properties
are influenced by the introduction of the porous array. In fact,
if the $H_{c2\bot}$ second derivative versus the temperature is
plotted we can see that it changes its sign from positive to
negative at $H \approx$ 1.16 Tesla. This field value is very close
to the nominal first matching field that we expect for the porous
Si template, $H_{1}\approx$ 1.30 Tesla, assuming a square porous
array. This change in concavity was already reported in a previous
study on the same kind of samples, and it was ascribed to the
formation of a commensurate vortex structure \cite{Trezza-JAP08}.
From the measured value of $H_{1}$ it follows that the period of
the porous template is $a$ = 42 nm. In the following we will
identify $a_{0}$ $\equiv$ 42 nm. In Fig. \ref{Fig.1}(b) is
reported the $H_{c2\bot}(T)$ curve for the Nb reference film of
the same thickness deposited on the non-porous template. As
expected the $H_{c2\bot}(T)$ behavior is linear over the all
temperature range and the $H_{c2\bot}$ second derivative versus
temperature does not present any peculiarity except for a shallow
peak near $T_c$. In the inset of Fig. \ref{Fig.1}(b), for sake of
comparison, the $dH_{c2\bot}^{2}$/$dT^{2}$ versus the reduced
temperature is reported for both the Nb films, in order to point
out the difference in their magnitude. A fit to the data close to
$T_{c}$ with the expression for $H_{c2\bot}(T)$ reported above,
yields a value of the Ginzburg-Landau coherence length at $T$ = 0,
$\xi_{0\parallel}$ = 9.1 nm and $\xi_{0\parallel}$ = 9.5 nm,
resulting in a superconducting coherence length $\xi_{S}$ = 5.8 nm
and $\xi_{S}$ = 6.0 nm, for the Nb porous sample and the Nb
reference film, respectively. The values of $\xi_{0\parallel}$ are
significantly smaller than the BCS coherence length of Nb,
$\xi_{0}$ = 39 nm \cite{Buckel}, indicating that our films are in
dirty limit regime with an electron mean free path of $\textit{l}$
= 1.38 $\xi_{0\parallel}^{2}$ / $\xi_{0}$ $\approx$ 3 nm
\cite{Schmidt}. Since the film dimensions in the $xy$ plane are
larger than $\xi_{\parallel}(T)$, the expression for
$H_{c2\bot}(T)$, reported above, is verified in the whole
temperature range. The Ginzburg-Landau parameter, $\kappa$ =
$\lambda$(0)/$\xi_{0\parallel}$, can be estimated using the
expression $\kappa$ = 0.72$\lambda_{L}$/$\textit{l}$ = 9.6, where
$\lambda_{L}$ = 39 nm is the London penetration depth of Nb
\cite{Buckel}. Ratios of $\xi_{0\parallel}$/$\textit{a} \approx$
0.2 and $\lambda$(0)/$\textit{a} \approx$ 2.1, measured for
$a_{0}$ = 42 nm, are larger than in previous works
\cite{Welp-PRB,Mosh06} on perforated Nb samples, and indicate that
we are in presence of individual vortex pinning \cite{Brandt}.
Moreover, the pore diameter, ${\O}$, in our PS template is
comparable with the vortex core dimension at $T$=0, ${\O} \approx
\xi_{0\parallel}$. This means that the saturation number, $n_{S}$
= $\frac{\O}{2\xi_{S}(T)}$, defined as the maximum number of
vortices that fits into a pore with diameter ${\O}$, is less or
equal to 1, so that each pore can trap only one fluxon
\cite{Mkrtchyan}. Subsequently multiquanta vortex lattice
\cite{Mosh98} cannot be observed in our system.\\
Now we move to a more careful inspection of the $R(H)$ curves of
the Nb porous film. This will lead to the observation of a
peculiar behavior of these transitions, whose analysis represents
the main subject of this work. In Figs. \ref{Fig.2}(a) and 2(b)
$R(H)$ curves obtained for two different values of the
temperature, $T$ = 3.490 K and $T$ = 3.531 K, respectively, are
presented.

\begin{figure}
\includegraphics[width=8.5cm]{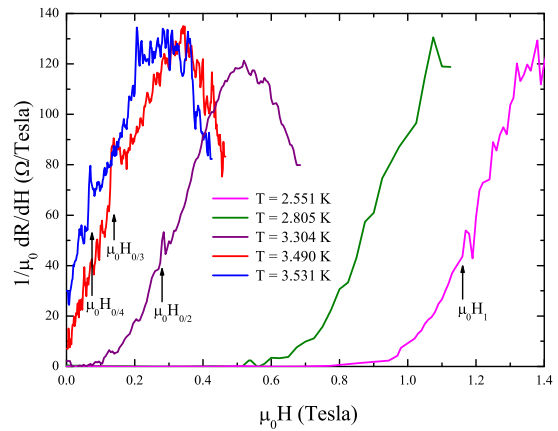} \caption{First derivatives, $dR$/$dH$, as a
function of the applied magnetic field at different temperatures.
The arrows indicate the field where the bump is present for each
temperature. (Color online)} \label{Fig.3}
\end{figure}

\noindent At first glance both the curves are rather smooth and do
not present any structures or enlargements due, for example, to
sample inhomogeneities. However if the dependence of the first
derivative $dR$/$dH$ versus the applied magnetic field is
analyzed, some distinct features can be observed. In particular in
both the curves a small local maximum is present at specific
values of the magnetic field. Let's focus on the position where
the bumps, as indicated by an arrow in Fig. \ref{Fig.2}, start to
develop. The bumps in the first derivative reflects the presence
of a small dip in the corresponding magnetic field dependence of
the resistance $R(H)$ at the same value of $H$. This effect was
ascribed to a pinning enhancement when the period of the vortex
structure is commensurate with the period of the antidots
\cite{Patel}. The bumps in the $dR$/$dH$ appear indeed in our
curves at values of the magnetic fields $H_{n}$ when the magnetic
flux threading each unit cell is equal to the flux quantum,
$\Phi_0$, or to fractional values of $\Phi_0$. In Fig.
\ref{Fig.2}(a), where the $R(H)$ measurement at $T$ = 3.490 K is
shown, the peculiarity in $dR$/$dH$ is, in fact, observed at
$H_{bump}\approx$ 0.126 Tesla. The period of the vortex lattice at
this field value is $a'$ = 128 nm, i.e. about three times the
interpore spacing of this analyzed sample, $a_{0}$ = 42 nm.
Consequently this field value corresponds to one-ninth of the
matching field $H_{1}$/9 $\approx$ 0.129 Tesla. Similarly, in Fig.
\ref{Fig.2}(b) where the $R(H)$ measurement at $T$ = 3.531 K is
shown, the bump in $dR$/$dH$ develops at $H_{bump}\approx$ 0.065
Tesla. The period of the vortex array at this field is then $a''$
= 178 nm, which is about four times the interpore spacing of this
sample. Consequently this field value corresponds to one-sixteenth
of the matching field $H_{1}$/16 $\approx$ 0.072 Tesla. An
additional bump structure is present at $H\approx$ 0.2 Tesla.
However, this field value does not correspond to any commensurate
vortex configuration (see discussion below) and does not survive
repeating the measurement in the same temperature range. Many
$R(H)$ measurements at different temperatures have been performed
and the behavior of all the corresponding $dR$/$dH$ curves has
been analyzed. A selection of these curves is reported in Fig.
\ref{Fig.3}. Some of them have been obtained by sweeping the field
upward and downward and no hysteresis has been detected.

For instance, the curves at $T$ = 2.551 K and $T$ = 3.304 K
present a bump at $H_{bump}$ = $H_{1}$ and $H_{bump}$ = $H_{1}$/4,
respectively. By comparison a curve with no bump, measured at
temperature $T$ = 2.805 K, is also shown. In all curves the fields
at which the bumps are observed are related to the first matching
field through the relation: $H$ = $H_{1}$/$n^{2}$ with $n$ =
1,...,4. The temperatures at which bumps are observed, the
corresponding fields and their values normalized to $H_{1}$, the
$\xi_{S}$ values, the vortex-vortex distances, $a$, and their
values normalized to $a_{0}$, are summarized in Table \ref{table}.

\begin{table}
\caption{Temperatures at which the bumps are observed,
corresponding fields and their values normalized to $H_{1}$,
$\xi_{S}$ values at that temperature, vortex-vortex distances,
$a_{k/l}$, and their values normalized to $a_{0}$ = 42 nm.}
\label{table}
\begin{center}
\begin{tabular}{cccccc}
$T$(K)  & $H_{bump}$(T) & $\frac{H_{bump}}{H_{1}}$ & $\xi_{S}$(nm) & $a_{k/l}$(nm) & $\frac{a_{k/l}}{a_{0}}$ \\
\hline
2.551 & 1.160  & 1 & 10.02  & 42.0  & 1.00  \\

3.304  & 0.275  & 1/4 & 15.66  & 87.0  &  2.07  \\

3.490  & 0.126  &  1/9 & 19.44  & 128.0  & 3.05  \\

3.531  & 0.065  & 1/16 & 20.76  & 178.0  & 4.24  \\
\end{tabular}
\end{center}
\end{table}

\begin{figure}
\includegraphics[width=7cm]{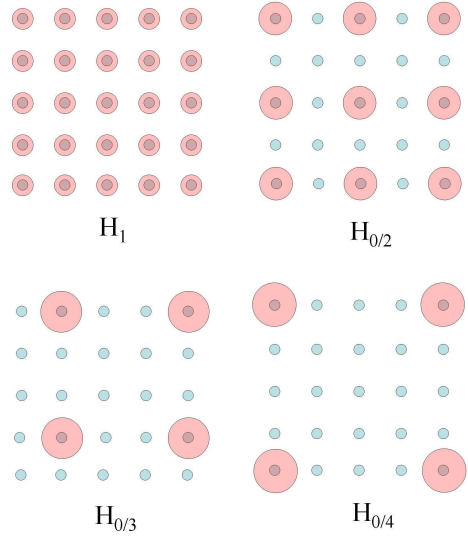} \caption{Vortex lattice configurations
occurring at the first matching field and its fractional values.
Increasing the temperature the vortices diameter and their
reciprocal distance increase, as reported in Table \ref{table}.
Blue circles represent the holes, pink ones represent the
vortices. (Color online)} \label{Fig.4}
\end{figure}

We argue that the presence of the observed bumps in the $dR$/$dH$
curves can be related to different vortex lattice arrangements
made possible by the lattice of holes. The specific vortex lattice
configurations occurring at the first matching field and at its
fractional values are shown in Fig. \ref{Fig.4}.

\begin{figure}
\includegraphics[width=8.8cm]{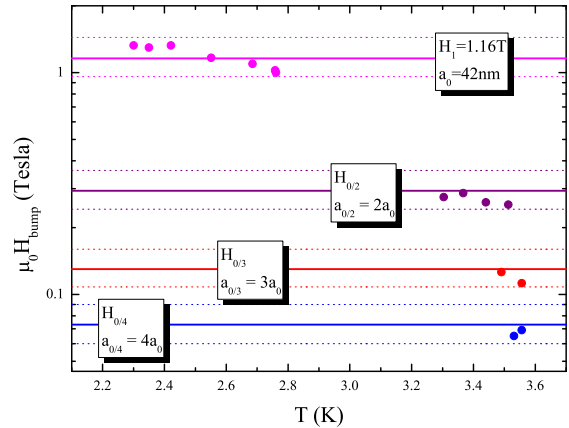} \caption{The points of coordinates
($H_{bump}$,$T$) identify the values of the fields and
temperatures at which bumps have been observed in the $dR$/$dH$
curves at fixed temperatures. The solid lines correspond to the
different matching field orders achieved with the interpore
spacing $a_{0}$ = 42 nm, while the dotted lines are obtained
forasmuch as the regularity of the pore distance is achieved
within the 10 percents of the average distance. (Color online)}
\label{Fig.5}
\end{figure}

\noindent In the case of $H_{bump}$/$H_{1}$ = 1 a commensurate
square vortex configuration is formed, where each pore is occupied
by a fluxon and the side of this square array is just $a_{0}$ = 42
nm. Increasing the temperature the vortices diameter
($\approx$2$\xi_{S}$) and their reciprocal distance increase, as
reported in Table \ref{table}. When $H_{bump}$/$H_{1}$ = 1/4, 1/9
and 1/16 a square vortex lattice is again obtained with $a$ = 87
nm, 128 nm and 178 nm, respectively. This means that the pores act
as an ordered template of strong pinning centers, which is able to
preserve the long range positional order of the flux lattice also
at low fields value, i.e. at higher vortex spacing. As already
pointed out the optimization of the vortex structures leads to the
formation of larger square flux lattices with respect to the
underlying artificial pinning array with the lattice constant $a$
exactly equal to $na_{0}$. The vortices tend to be placed as far
from each other as possible due to the repulsive interaction
between them and at the same time they want to follow the imposed
square potential induced by the antidots. This constraint gives
$a$ = $a_{0} \sqrt{l^{2}+k^{2}}$, where $l$ and $k$ are integer
numbers. Therefore, we should expect the fractional matching
fields at $H$ = $H_{k/l}$ = $\Phi_0/a^{2}$ =
$\Phi_0/[a_{0}^{2}(l^{2}+k^{2})]$ = $H_{1}/(l^{2}+k^{2})$
\cite{Mosh95}. We observed bumps at fractional matching fields
$H_{0/2}$, $H_{0/3}$ and $H_{0/4}$. The other bumps expected from
the equation above at fractional fields $H_{k/l}$ with $k \neq$ 0
have not been observed. All the fields values at which the bumps
in the $dR$/$dH$ appear are shown as points of coordinates
($H_{bump}$,$T$) in Fig. \ref{Fig.5}.

\noindent In this figure the solid lines correspond to the
matching fields of different order, as calculated assuming an
interpore spacing $a_{0}$ = 42 nm, through the formula
$H_1=\Phi_0/a_{0}^{2}$.  The dotted lines are obtained considering
a deviation from the corresponding mean interpore distance of the
order of 10\% \cite{Trezza-JAP08}. It is worth noticing that all
the data fall into the range theoretically estimated, suggesting
that the observed peculiarities in the $R(H)$ curves can be indeed
ascribed to commensurability effect between the porous structure
of the Nb film and the vortex lattice. The distribution of the
experimental points is consistent with the observation that a
certain temperature dependence of the matching effect can be found
for the case of short-range ordered templates \cite{Ziemann}. We
would also point out that the effect is observable in our sample
only up to $H$ = $H_{1}$, due to the very high value of the first
matching field. The second matching field in fact is $H_{2}$ =
2$H_{1}$ = 2.32 Tesla. From a linear extrapolation of the
$H_{c2\bot}$ curve, it follows that in order to see at this field
a bump in the $dR$/$dH$ we should measure a $R(H)$ curve at $T$ =
1.73 K, temperature which cannot be reached in our experimental
setup. All the field values reported above have been calculated
assuming a square lattice. The measured field values do not match
with the ones calculated if a triangular array for the pores is
considered. In fact, at $T$ = 3.490 K (see Fig. \ref{Fig.2}(a))
the structure in the $dR$/$dH$ curve for a triangular lattice
would have been observed at a field $2/\sqrt{3}$ times higher than
$H_{0/3}$ = 0.126 Tesla, where no peculiar feature has been
detected. This supports our assumption of considering a square
lattice of holes in our system.

\section{Conclusions}

Matching effects have been reported for Nb thin film grown on
porous silicon. Due to the extremely reduced values of the
interpore distance the effect is present at fields values higher
than 1 Tesla and down to reduced temperatures as low as t $\simeq$
0.52. The commensurability manifests both in the ($H$,$T$) phase
diagram and in the $R(H)$ transitions. The latter in particular
reveal the formation of fractional matching states. As it was
argued in many works the vortex configuration at fractional
matching fields are characterized by striking domain structure and
associated grain boundaries \cite{Field02,Mosh03}. The presence of
multiple degenerate states with domain formation at the fractional
field, directly observed with scanning Hall probe microscopy
\cite{Field02}, seems to be high probable in our films. The
reduced regularity of our templates, in fact, could be compensated
by the formation of domain walls of different complexity. The
particular domain configuration is of course a matter of energy
balance between the cost in energy for the wall formation and the
energy gain due to the vortex pinning.





\end{document}